\documentclass{article}

\usepackage{arxiv}

\usepackage[utf8]{inputenc} 
\usepackage[T1]{fontenc}    
\usepackage[bookmarks=false]{hyperref}
\usepackage{url}            
\usepackage{booktabs}       
\usepackage{amsfonts}       
\usepackage{nicefrac}       
\usepackage{microtype}      
\usepackage{lipsum}
\usepackage{graphicx,xcolor}
\usepackage{subfigure}
\usepackage{amssymb,amsfonts,amsbsy,amsmath}
\usepackage{comment}
\usepackage{epstopdf} 
\usepackage[export]{adjustbox} 
\usepackage{adjustbox} 

\DeclareMathOperator*{\argmax}{arg\,max}
\usepackage{ifpdf}
\usepackage{float}

\title{A novel spike-and-wave automatic detection \\in EEG signals}

\author{
  Antonio Quintero-Rinc\'on, Valeria Muro, Carlos D'Giano \\
  Fundaci\'on Lucha contra las Enfermedades Neurol\'ogicas Infantiles (FLENI)\\
  Buenos Aires, Argentina \\
  \texttt{tonioquintero@ieee.org} \\
   \And
 Jorge Prendes, Hadj Batatia \\
  University of Toulouse, IRIT - INPT\\
	Toulouse, France}

\begin{document}
\maketitle

\begin{abstract}
Spike-and-wave discharge (SWD) pattern classification in electroencephalography (EEG) signals is a key problem in signal processing. It is particularly important to develop a SWD automatic detection method in long-term EEG recordings since the task of marking the patters manually is time consuming, difficult and error-prone.
This paper presents a new detection method with low computational complexity that can be easily trained, if standard medical protocols are respected. 
The detection procedure is as follows:  
First, each EEG signal is divided into several time segments and for each time segment, the Morlet 1-D decomposition is applied.
Then three parameters are extracted from the wavelet coefficients of each segment: \emph{scale} (using a generalized Gaussian statistical model), \emph{variance} and \emph{median}. This is followed by a $k$-nearest neighbors ($k$-NN) classifier to detect the spike-and-wave pattern in each EEG channel from these three parameters. A total of 106 spike-and-wave and 106 non-spike-and-wave were used for training, while 69 new annotated EEG segments from six subjects were used for classification. In these circumstances, the proposed methodology achieved 100\% accuracy. These results generate new research opportunities for the underlying causes of the so called \emph{absence epilepsy} in long-term EEG recordings.
\end{abstract}

\keywords{Spike-and-wave \and Generalized Gaussian distribution \and Electroencephalography \and Morlet wavelet \and $k$-nearest neighbors classifier \and Epilepsy}

\section{Introduction}
Epilepsy is a chronic neurological disorder that affects patients with recurrent seizures. Seizures are characterized by the excessive electrical discharges in neurons, their  waveform is known as a spike. A spike is characterized by short bursts of high amplitude, synchronized and multi-phasic activity, where polarity changes occur several times. Spikes manifest themselves at or around the epileptic focus and stand out from the background EEG activity. Electroencephalography (EEG) is currently the main technique to record electrical activity in the brain. Specific Neurologists, trained in EEG, are able to properly determine an epilepsy diagnosis analyzing the different types of spikes in the so called \emph{rhythmic activity} of the brain. 

Automatic methods for detecting epileptic events in EEG signals greatly exceed visual inspection. These methods focus on interictal spikes \cite{Bergstrom2013,Bhuyan2017}, seizure onset detection \cite{QuinteroRincon2016a} or waveforms epileptic patterns \cite{Gajic2014,Navakatikyan2006}. There is a wide variety of EEG signal processing features such as spatial-temporal analysis \cite{Ossadtchi2010}, frequency-temporal analysis \cite{Wilson2002}, wavelet decomposition \cite{Bergstrom2013}, spectrogram \cite{VanHese2009,Pearce2014}, Hilbert transform \cite{Kamath2014}, neural networks \cite{Puspita2017}, Hurst exponent \cite{Indiradevi2009} or by statistical model \cite{QuinteroRincon2017b}. In general these features are used with classifiers based on machine learning techniques with high performance such as Support Vector Machine \cite{Siuly2015}, logistic regression \cite{Subasi2005}, decision trees \cite{QuinteroRincon2017c}, k- Nearest Neighbor \cite{DiGesu2009,Rezaee2016} or Random Forest \cite{Donos2015}. 

Spike-and-wave discharge (SWD) is a generalized EEG discharge pattern whose waveform has a regular and symmetric morphology.  This morphology can be mathematically described by a Morlet wavelet transform that generates a time-frequency representation of the EEG signal \cite{Subasi2005,Xanthopoulos2009,Sitnikova2009,Richard2015}. The \emph{spike} component of a SWD is associated with neuronal firing and the \emph{wave} component is associated with neuronal inhibition or hyperpolarization of neurons \cite{Pollen1964}.   SWD is widely used in mice studies \cite{VanHese2009,Ovchinnikov2010,Bergstrom2013,Rodgers2015}  nonetheless, human testing is limited. Mice have a predisposition for generalized SWDs at 7-12 Hz \cite{Pearce2014} and typically have spontaneous absence-seizure-like-events. The the presence of an intact cortex, thalamus and their interconnections is necessary to record them \cite{Blumenfeld2005,Avoli2012}. Some recent works in humans, using feature extraction to estimate the SWD pattern coupled with  machine learning classification, have been proposed in \cite{QuinteroRincon2017b} with the t-location-scale distribution as feature extraction and a $k$-NN classifier, in \cite{QuinteroRincon2017c} through cross-correlation coupled with decision trees and in \cite{Puspitaa2017} using Bayesian classification from the Walsh transformation which is used to define the SWD morphological characteristic. In other works a Hilbert-Huang transform is estimated to assess the characteristics of time-frequency energy distributions \cite{Zhu2015} or by using complex network of neuronal oscillators \cite{Medvedeva2018} or through the different parameters such as: variance, sum of wave amplitudes, slope of wave and mobility of the waveform \cite{Olejarczyk2009}.

The remaining of this paper is structured as follows. Section \ref{sec:method} presents the proposed method and explains the Morlet wavelet, the generalized Gaussian distribution (GGD) model and the SWD detection by $k$-NN classifier based on the extracted parameters. Experimental results are reported in section \ref{sec:results} where the classifier vector is composed by the scale parameter of the GGD and the variance and median of the wavelet coefficients from EEG data. Lastly, the conclusion, remarks, and future perspectives are presented in Section \ref{sec:conclusion}.

\section{Proposed method}
\label{sec:method}
This section presents a new statistical method to detect spike-and-wave discharges (SWD) in EEG signals. The methodology is computationally very efficient, suitable for real-time automation, and can be used to perform the spike-and-wave detection on-line. First, the database used in this study will be introduced, then the methodology will be explained.

\subsection{Database}
\label{subsec:dB}
\begin{figure}[!t]
	\centering
	\subfigure[SWD and Morlet Wavelet]{\includegraphics[width=100mm]{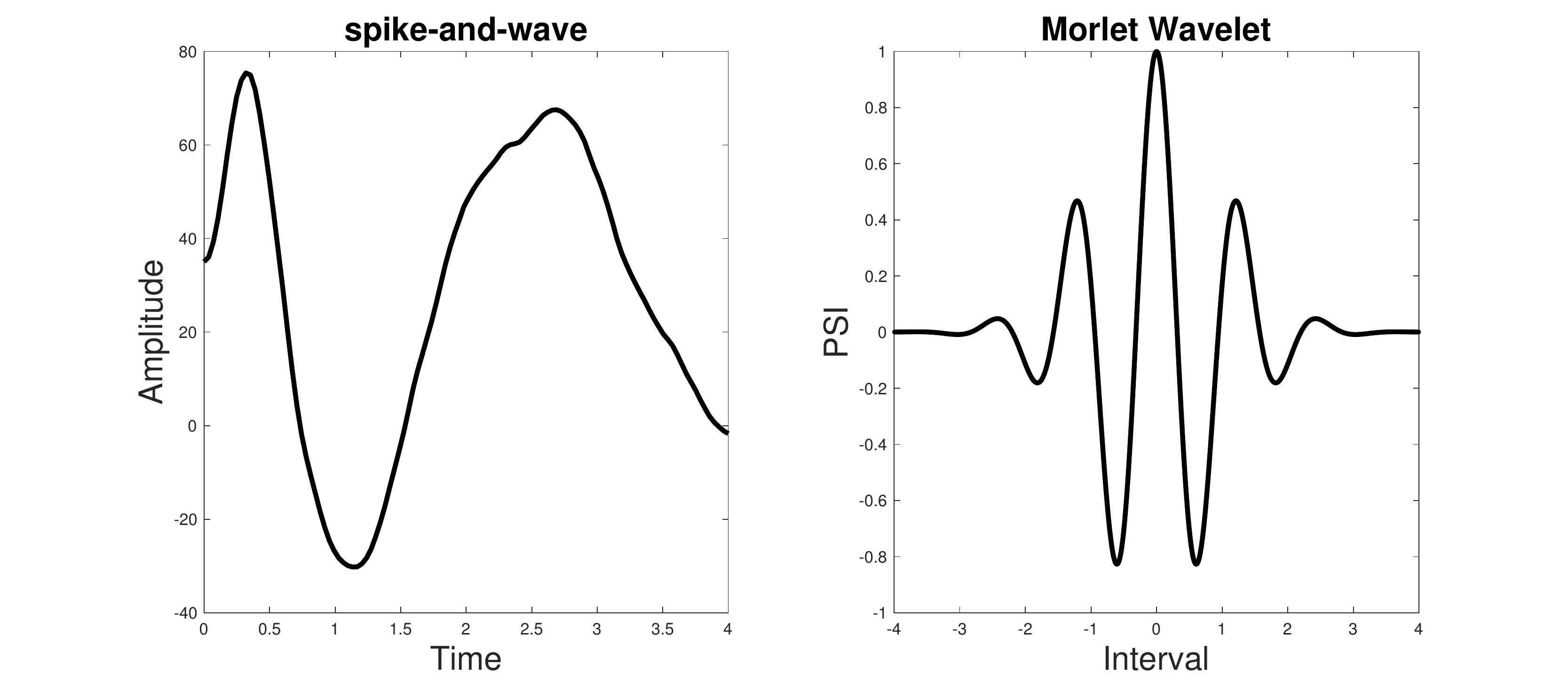}}
	\subfigure[EEG example]{\includegraphics[width=100mm]{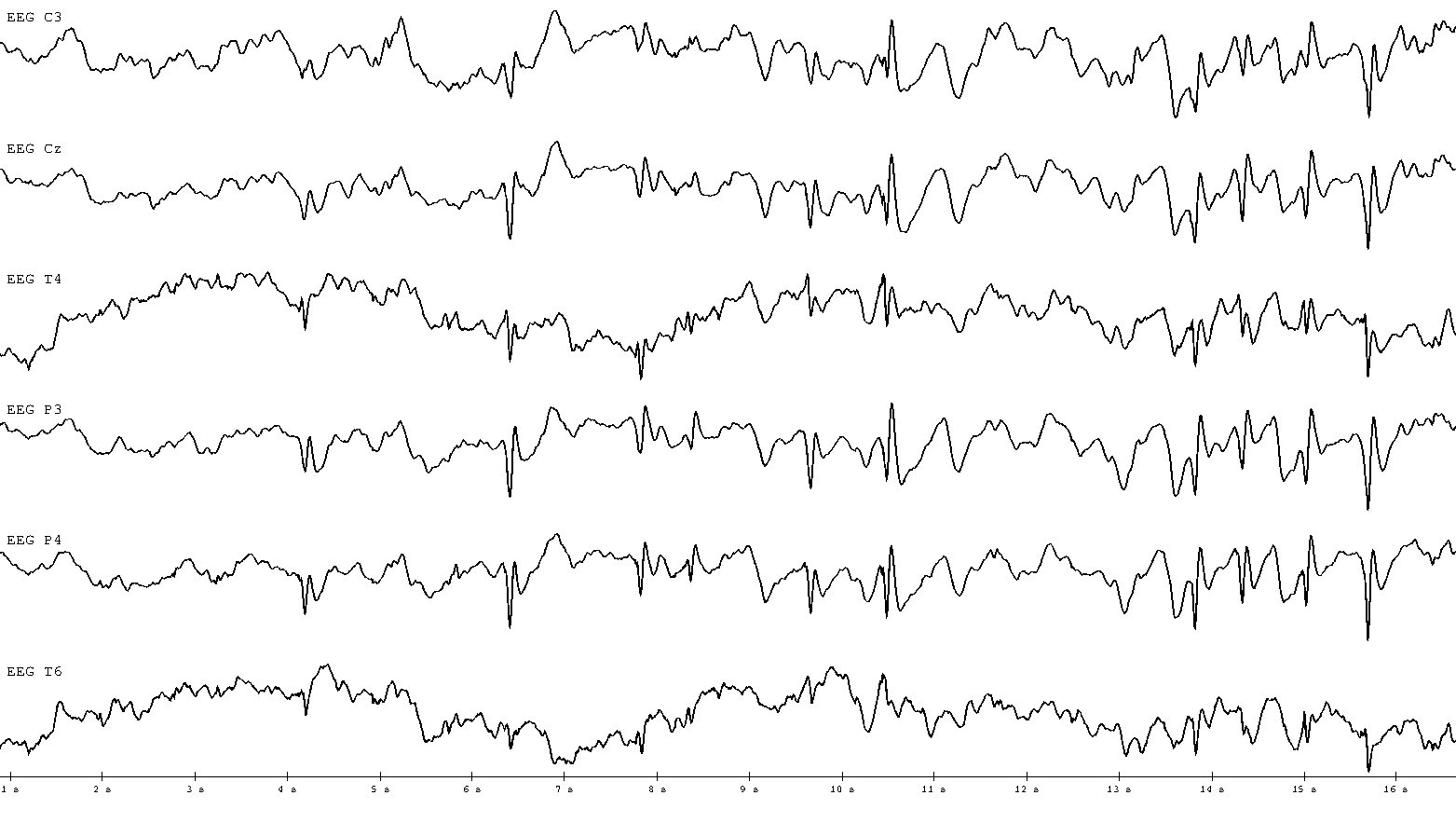}}
	\caption{(a) SWD signal and Morlet wavelet respectively, we can see the symmetric and regular morphology in both signals. (b) Example of 6 channels of one monopolar raw EEG, we can see several SWDs in all channels.}
	\label{fig:swdmorlet}
\end{figure}

A database with 212 monopolar 256Hz signals was created for off-line training of the classifier: 106
SWD signals and 106 non-SWD signals, measured from six patients from \emph{Fundaci\'on Lucha contra las Enfermedades Neurol\'ogicas Infantiles} (FLENI).
 The SWD signals have different times and waveforms but their morphology is preserved, while the non-SWD signals have normal waveforms. See Figure \ref{fig:swdmorlet} for an example of a typical SWD signal.

Analyzing each SWD in the frequency domain, it was observed that they are restricted to a narrow frequency band between 1-3 Hz. Each EEG was acquired with a 22-channels array using the standard $10/20$ system through the following channels:  Fp1, Fp2, F7, F3, Fz, F4, F8, T3, C3, Cz, C4, T4, T5, P3, Pz, P4, T6, O1, O2, Oz, FT10 and FT9. 

All new segments to analyze contain different spike-and-waves events. Their onset and duration time has been labeled by an expert neurologist. Here we used the expert annotations to extract a short epoch from each recording such that it is focused on the spike-and-wave in long-time EEG signals (the epochs used  have a duration of the order of 1 minute). 

It should be noted that, for each new patient to analyze, ten new SWD are selected to be part of the database. This permits to be Patient-specific seizure detection.

\subsection{Methodology}
\label{subsec:methodology}
Let $\boldsymbol X \in \mathbb{R}^{N\times M}$ denote the matrix gathering $M$ EEG signals $\boldsymbol{x}_{m} \in \mathbb{R}^{N\times 1}$ measured simultaneously on different channels and at $N$ discrete time instants.
The proposed method is composed of four stages. The first stage splits the original signal $\boldsymbol X$ in several segments of 2 seconds with 1 second overlap using a rectangular sliding window so that
\begin{align}
	\boldsymbol X_{t} &= \boldsymbol \Omega_{t} \boldsymbol X\\
	\boldsymbol \Omega_{t} &= \left[\boldsymbol0^{L\times tL}, \boldsymbol{I}^{L\times L},\boldsymbol0^{L\times N-tL-L}\right]
\end{align}
where $\boldsymbol 0^{N\times M} \in \mathbb R^{N\times M}$ is the null matrix, $\boldsymbol I^{N\times N} \in \mathbb R^{N\times N}$ is the identity matrix and $L$ is the number of measurement obtained in 2 seconds. The second stage consists of representing each segment $\boldsymbol X_{t}$ using a time-frequency Morlet decomposition. The purpose of this decomposition is to evaluate the energy distribution throughout SWD frequency band, which is restricted to a narrow 1-3 Hz frequency window. Then a time/scale relationship is applied in order to find the Morlet wavelet coefficients. In the third stage, the statistical distribution of the Morlet wavelet coefficients are represented by using a zero-mean generalized Gaussian distribution. Each wavelet coefficient is summarized by estimating the statistical parameters \emph{scale ($\varsigma$)} and \emph{shape ($\tau$)} of the generalized Gaussian distribution similar to our previous works \cite{QuinteroRincon2016a,QuinteroRincon2016b,QuinteroRincon2017a,QuinteroRincon2018a,QuinteroRincon2018b}. In these jobs we found that the scale parameter $\varsigma$ closely relates to the variability
of the brain activity and is therefore a good descriptor for performing seizure detection. Therefore the scale parameter (which depends on the shape parameter) is enough to detect the seizure in EEG signals.

\subsubsection{Morlet Wavelet}
The Continuous Wavelet Transform is given by
\begin{align}
	W_{f}(t,a,b) &= \int_{-\infty}^{\infty} \boldsymbol X_{t} \; \psi^{*}_{a,b}(t) \text{dt} 
	\label{eq:wav} \\
	\psi^{*}_{a,b}(t) &=\frac{1}{\sqrt{a}}\; \psi \left(\frac{t-b}{a}\right)
	\label{eq:wavmother}\\
	\psi(t) &= \exp^{- \frac{t^{2}}{2}}cos(5t)
	\label{eq:morlet}
\end{align}
where $a$ is the scaling parameter, $b$ is the shifting parameter, \eqref{eq:wavmother} is the mother wavelet function, $(*)$ denotes the complex conjugate operation and \eqref{eq:morlet} is the analytic expression of the Morlet wavelet \cite{Ahuja2005}. In order to associate the Morlet wavelet as a purely periodic signal of frequency $F_{c}$, we use the relationship between scale and frequency 
\begin{align}
	F_{a} = \frac{F_{c}}{\alpha \Delta}
	\label{eq:scal2frq}
\end{align}
where $\alpha$ is the scale, $\Delta$ is the sampling period, $F_{c}$ is the center frequency of Morlet wavelet in Hz and $F_{a}$ is the pseudo-frequency corresponding to the scale $a$ in Hz. The center frequency-based approximation captures the main wavelet oscillations. Therefore, the center frequency is a convenient and simple characterization of the dominant frequency of the wavelet \cite{Abry1997}. Note that the wavelet scale is estimated according the 1-3 Hz restricted narrow frequency of the SWD database.

\subsubsection{Generalized Gaussian distribution}
\label{ssec:ggd}

The univariate generalized Gaussian distribution (GGD) is a flexible statistical model for one-dimensional signals that has found numerous applications in science and engineering.  The distribution of the Morlet wavelet coefficients $\boldsymbol C_{t}$ can be represented by using a zero-mean GGD statistical model \cite{Do2002} with probability density function (PDF) given by
\begin{align}
	\label{eq:PDFGGD2}
	f_{\textnormal{GGD}}(x;\varsigma,\tau) = \frac{\tau}{2\varsigma\Gamma(\tau^{-1})} \exp\left(-\left|\frac{x}{\varsigma}\right|^\tau\right)
\end{align}
where $\varsigma \in \mathbb{R}^+$ is a scale parameter, $\tau \in \mathbb{R}^+$ is a shape parameter that controls the shape of the density tail and $\Gamma\left(\cdot\right)$ is the Gamma function. From \eqref{eq:PDFGGD2}, the statistical properties of a wavelet coefficient matrix $\boldsymbol C_{t}$ can be summarized with the maximum likelihood parameter vector $\boldsymbol\Theta_{\boldsymbol C_{t}}$:
\begin{align}
	\boldsymbol\Theta_{\boldsymbol C_{t}}
	&= \left[\varsigma_{t},\tau_t\right]^{T} = \argmax_{\left[\varsigma,\tau\right]^{T}} f_\textnormal{GGD}(\boldsymbol C_{t};\varsigma,\tau)
\end{align}

For more detais about the GGD parameters, we refer the reader to our previous works  \cite{QuinteroRincon2016a,QuinteroRincon2016b,QuinteroRincon2017a,QuinteroRincon2018a, QuinteroRincon2018b}.

\subsubsection{Spike-and-wave detection using a k-nearest neighbors classifier}
Consider a classification into two possible classes $c=0$ and $c=1$, then for a feature vector $\boldsymbol \Theta_{\boldsymbol C_{t}}$ each class is given by
\begin{align}
	\nonumber
	\rho\left(\boldsymbol\Theta_{\boldsymbol C_{t}}|c=0\right) &= \frac{1}{N_{0}}\sum_{n \in \textnormal{class 0}}\mathcal{N}\left(\boldsymbol\Theta_{\boldsymbol C_{t}}|\boldsymbol\Theta_{\boldsymbol C_{t}}^n,\sigma^{2}   \boldsymbol I\right) \\
	\label{eq:class1}
	&=\frac{1}{N_{0}\left(2\pi\sigma^{2}\right)^{D/2}}\sum_{n \in \textnormal{class 0}} \exp^{-\frac{\left(\boldsymbol\Theta_{\boldsymbol C_{t}}-\boldsymbol\Theta_{\boldsymbol C_{t}}^n\right)^{2}}{2\sigma^{2}}} \\
	\nonumber
	\rho\left(\boldsymbol\Theta_{\boldsymbol C_{t}}|c=1\right) &= \frac{1}{N_{1}}\sum_{n \in \textnormal{class 1}}\mathcal{N}\left(\boldsymbol\Theta_{\boldsymbol C_{t}}|\boldsymbol\Theta_{\boldsymbol C_{t}}^n,\sigma^{2}   \boldsymbol I\right) \\
	\label{eq:class2}
	&=\frac{1}{N_{1}\left(2\pi\sigma^{2}\right)^{D/2}}\sum_{n \in \textnormal{class 1}} \exp^{-\frac{\left(\boldsymbol\Theta_{\boldsymbol C_{t}}-\boldsymbol\Theta_{\boldsymbol C_{t}}^n\right)^{2}}{2\sigma^{2}}}
\end{align}
where $D$ is the dimention of a datapoint $\boldsymbol\Theta_{\boldsymbol C_{t}}$, $N_{0}$ or $N_{1}$ are the number of train points of class $0$ or class $1$ respectively and $\sigma^{2}$ is the variance. Using Bayes rule to classify a new datapoint $\boldsymbol\Theta_{\boldsymbol C_{t}}^{*}$ in each class the following equation is obtained
\begin{align}
	\rho\left(c=0|\boldsymbol\Theta_{\boldsymbol C_{t}}^{*}\right) =\frac{\rho\left(\boldsymbol\Theta_{\boldsymbol C_{t}}^{*}|c=0\right)\rho\left(c=0\right)}{\rho\left(\boldsymbol\Theta_{\boldsymbol C_{t}}^{*}|c=0\right)\rho\left(c=0\right)+\rho\left(\boldsymbol\Theta_{\boldsymbol C_{t}}^{*}|c=1\right)\rho\left(c=1\right)}
	\label{eq:bayes} 
\end{align}
The Maximum Likelihood setting of $\rho(c = 0)$ is $N_{0}/(N_{0}+N_{1})$, and $\rho(c = 1) = N_{1}/(N_{0}+N_{1})$. An analogous expression to equation \eqref{eq:bayes} can be obtained for $\rho\left(c=1|\boldsymbol\Theta_{\boldsymbol C_{t}}^{*}\right)$. To determine which class is most likely, the ratio between both expression is used, which simplifies as follows:
\begin{align}
	\frac{\rho\left(c=0|\boldsymbol\Theta_{\boldsymbol C_{t}}^{*}\right)}{\rho\left(c=1|\boldsymbol\Theta_{\boldsymbol C_{t}}^{*}\right)} = \frac{\rho\left(\boldsymbol\Theta_{\boldsymbol C_{t}}^{*}|c=0\right)\rho\left(c=0\right)}{\rho\left(\boldsymbol\Theta_{\boldsymbol C_{t}}^{*}|c=1\right)\rho\left(c=1\right)}
	\label{eq:ratio}
\end{align}
If this ratio is greater than one, $\boldsymbol\Theta_{\boldsymbol C_{t}}^{*}$ is classified as $c=0$, otherwise it is calssified as $c=1$. It is important to note that in the case where $\sigma^{2}$ is very small in \eqref{eq:ratio}, then both the numerator as denominator will be dominated by the term for which the datapoint $\boldsymbol\Theta_{\boldsymbol C_{t}}^{n_{0}}$ in class $0$ or  $\boldsymbol\Theta_{\boldsymbol C_{t}}^{n_{1}}$ in class $1$ are closest to the point $\boldsymbol\Theta_{\boldsymbol C_{t}}^{*}$ respectively, such that
\begin{align}
	\nonumber
	\frac{\rho\left(c=0|\boldsymbol\Theta_{\boldsymbol C_{t}}^{*}\right)}{\rho\left(c=1|\boldsymbol\Theta_{\boldsymbol C_{t}}^{*}\right)} &=  \frac{exp^{-\frac{\left(\boldsymbol\Theta_{\boldsymbol C_{t}}^{*} - \boldsymbol\Theta_{\boldsymbol C_{t}}^{n_{0}}\right)^{2}}{2\sigma^{2}}}\rho\left(c=0\right)/N_{0}}{exp^{-\frac{\left(\boldsymbol\Theta_{\boldsymbol C_{t}}^{*} - \boldsymbol\Theta_{\boldsymbol C_{t}}^{n_{1}}\right)^{2}}{2\sigma^{2}}}\rho\left(c=1\right)/{N_{1}}} \\
	&= \frac{exp^{-\frac{\left(\boldsymbol\Theta_{\boldsymbol C_{t}}^{*} - \boldsymbol\Theta_{\boldsymbol C_{t}}^{n_{0}}\right)^{2}}{2\sigma^{2}}}}{exp^{-\frac{\left(\boldsymbol\Theta_{\boldsymbol C_{t}}^{*} - \boldsymbol\Theta_{\boldsymbol C_{t}}^{n_{1}}\right)^{2}}{2\sigma^{2}}}}
	\label{eq:ratiored}
\end{align}
On the limit $\sigma^{2} \to 0$, $\boldsymbol\Theta_{\boldsymbol C_{t}}^{*}$ is classified as class $0$ if $\boldsymbol\Theta_{\boldsymbol C_{t}}^{*}$ has a point in the class $0$ data which
is closer than the closest point in the class $1$ data. The nearest (single) neighbor method is therefore recovered as the limiting case of a probabilistic generative model. We refer the reader to
\cite{Bishop2006,BayesMachineLearning2012} for a comprehensive treatment of the mathematical properties of $k$-nearest neighbors.

\section{Results}
\label{sec:results}

In the training stage the annotated database introduced in Section \ref{subsec:dB} was utilized. These 212 monopolar signals were trained off-line using a $k$-nearest neighbors on a modified vector $[\varsigma, \sigma^{2}, \widetilde{x}] \in \mathbb{R}^3$ collecting the parameters associated with the Morlet wavelet coefficients for each 2-second segment with 1 second overlap, where $\varsigma$ is the scale parameters of the generalized Gaussian distribution, $\sigma^{2}$ is the variance parameter and $\widetilde{x}$ is the median parameter of the feature vector $\boldsymbol\Theta_{\boldsymbol C_{t}}$.  

\begin{figure}[H]
	\centering
	\subfigure[sigma ($\varsigma$) vs variance ($\sigma^{2}$)]{\includegraphics[width=100mm]{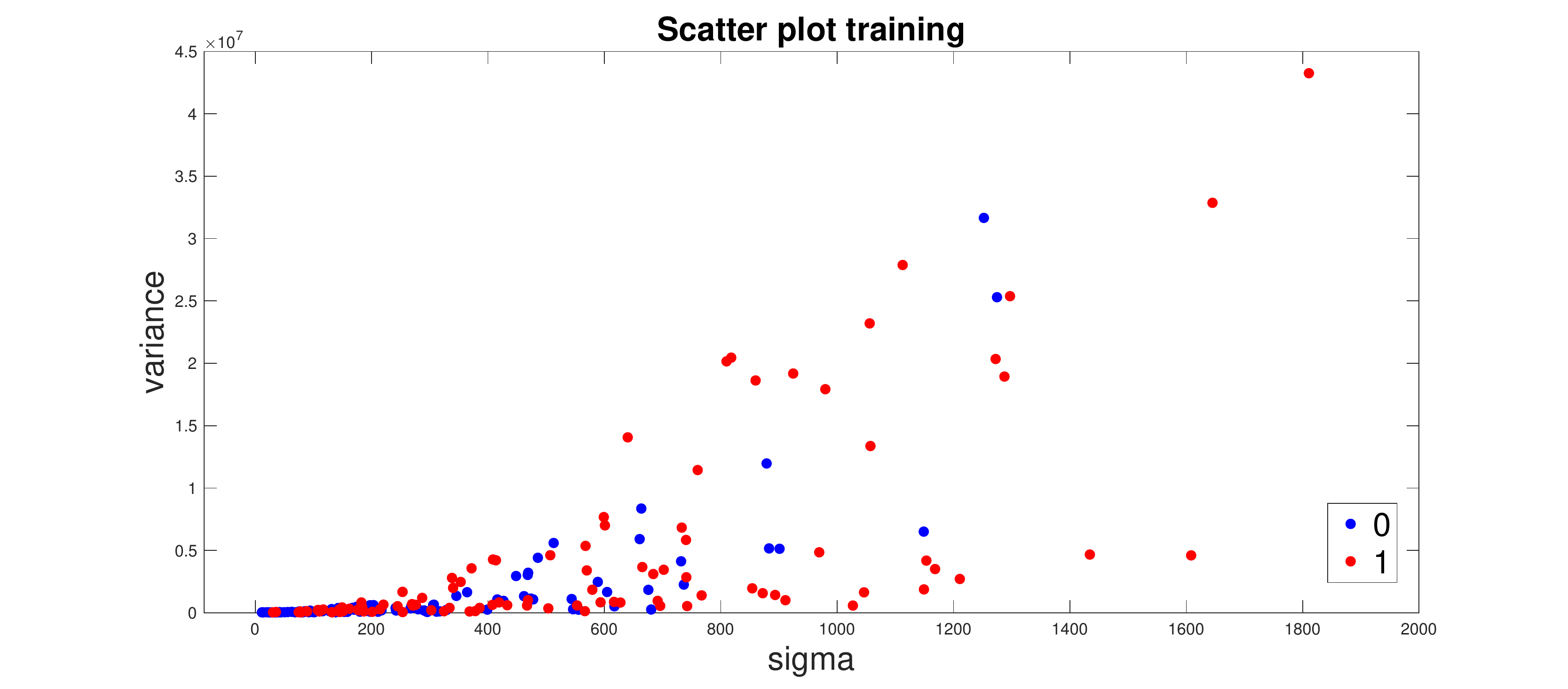}}
	\subfigure[sigma ($\varsigma$) vs median ($\widetilde{x}$)]{\includegraphics[width=100mm]{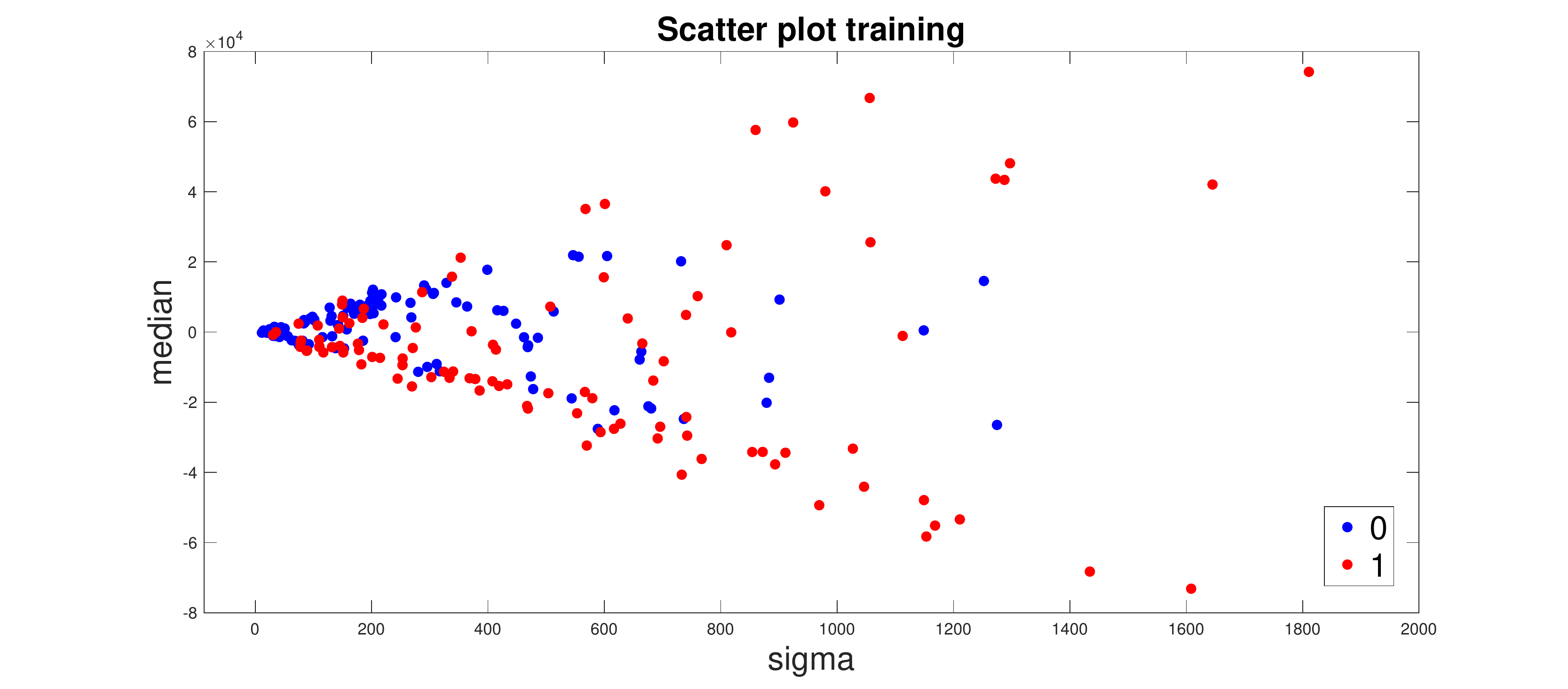}}
	\subfigure[variance ($\sigma^{2}$) vs median ($\widetilde{x}$)]{\includegraphics[width=100mm]{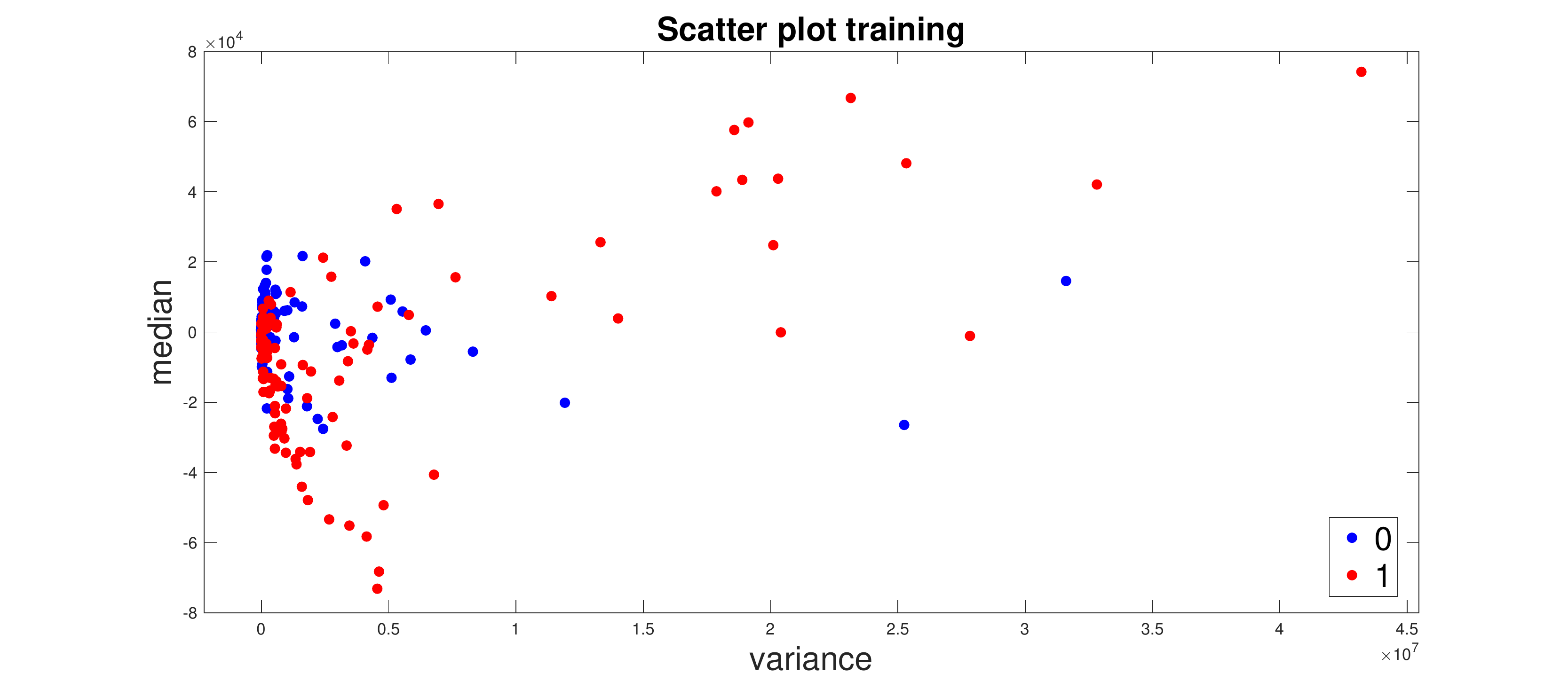}}
	\caption{Scatter plots of the off-line training classification in database signals, for $\varsigma$, $\sigma^{2}$ and $\widetilde{x}$ parameters for spike-and-waves events (SWD = class 1 = red dots) and non-spike-and-waves events (non-SWD = class 0 = blue dots), showing the data dispersion of the proposed approach. In (a) Scale parameter ($\varsigma$) vs variance ($\sigma^{2}$). For class 1 or SWD, we can see the direct relationship between the variance and sigma, both grow proportionally, while for class 0 or non-SWD  both sigma and variance remain in a range of values.
		(b) Scale parameter ($\varsigma$) vs median ($\widetilde{x}$). As sigma grows, median increases and decreases for both SWD and non-SWD, but is larger for SWD.
		(c) variance ($\sigma^{2}$) vs median ($\widetilde{x}$). As variance grows, median increases and decreases for SWD, while for non-SWD, remains in a small range.}
	\label{fig:Training}
\end{figure}

Table \ref{tab:range} contains the different bounds for each parameter, note that both minimum and maximum values are large for $\varsigma$, $\sigma^{2}$ and $\widetilde{x}$, when SWD and non-SWD signals are compared. This observation suggests that a threshold could be implemented to detect SWD patterns as a clear discrimination exists between spike-and-wave with respect to non-spike-and-wave. To illustrate that mentioned above, Figure \ref{fig:Training} shows scatter plots for the tree parameters in the next couples: 
\begin{enumerate}
	\item Scale parameter ($\varsigma$) vs variance ($\sigma^{2}$): For class 1 (SWD), a direct relationship between the variance and sigma, where both parameters grow proportionally, can be detected. For class 0 (non-SWD), both sigma and variance remain in a limited range of values.
	\item Scale parameter ($\varsigma$) vs median ($\widetilde{x}$): As sigma grows, median increases and decreases for both SWD and non-SWD, but is larger for SWD. A cone shaped pattern can be identified.
	\item Variance ($\sigma^{2}$) vs median ($\widetilde{x}$): As the variance grows, the median increases and decreases for SWD, while for non-SWD it remains in a small range (cluster).
\end{enumerate}

\vspace*{-1em}
\begin{table}[H]
	\caption{Range of values for sigma ($\varsigma$), variance ($\sigma^{2}$) and median ($\widetilde{x}$) parameters for class 0 or non-spike-and-wave and for class 1 or spike-and-wave.}
	\centering
	\begin{tabular}{||c|| c||c||c||}
		\hline \hline
		Metric    &	 Sigma ($\varsigma$) &	 Variance ($\sigma^{2}$) & Median ($\widetilde{x}$) \\
		\hline \hline
		Class 0 & [$12$, $1300$] & [$950$, $32\times10^6$] & [$-28\times10^3$, $22\times10^3$] \\
		\hline \hline
		Class 1 & [$31$, $1800$] & [$2800$, $43\times10^6$] & [$-73\times10^3$, $74\times10^3$] \\
		\hline \hline 	
	\end{tabular}
	\label{tab:range}
\end{table}

The performance of the $k$-nearest neighbors classification method using 10 neighbors with 3 predictors: $\varsigma$, $\sigma^{2}$ and $\widetilde{x}$ was evaluated using a dataset consisting of 69 new annotated measurement.  The new 69 annotated dataset correspond to 69 segments extracted from new six EEG signals of different subjects from the \emph{Fundaci\'on Lucha contra las Enfermedades Neurol\'ogicas Infantiles} (FLENI).
The assessment of the results was undertaken in terms of the overall accuracy of the classification. The classifier achieved a 100\% sensitivity (True Positive Rate) and specificity (True Negative Rate) for SWD detection.

\section{Conclusions}
\label{sec:conclusion}
This paper presents a new classification method to detect spikes-and-waves events in long-term EEG signals. The method proposed is based on the scale parameter of the generalized Gaussian distribution augmented with the variance and the median of the Morlet wavelet coefficients from EEG data and a $k$-nearest neighbors classification scheme that discriminates spike-and-wave from non-spike-and-wave events. 

The performance of the method was evaluated by training the algorithm with a real dataset containing 212 signals recordings with both spike-and-wave and non-spike-and-wave events. The classification performance was assessed utilizing 96 annotated segments and achieved a 100\% accuracy for spike-and-wave detection. This result sheds light on the potential for new research opportunities in the underlying causes of the so called \emph{absence epilepsy} in long-term EEG recordings. 

The main advantage is that the proposed algorithm can be implemented in real time and classifies, with high accuracy, the spike-and-wave pattern in epileptic signals. It is considered that these excellent results are due to the fact that, for each new patient analyzed, ten new SWD patterns are selected to be part of the database before training. Once the entire new database is trained, the prediction transforms into a patient-specfic seizure detection. The main limitation is defining the ideal sliding time-window and the overlap of segments due to the high dynamics of epileptic signals.

Future work will focus on other epileptic waveforms patterns as well as on an extensive evaluation of the proposed approach, comparison with other methods found in literature, implementing a medical-friendly interface with automatic count and increase the database of spike-and-waves in on-line EEG long-term signals detection.

\section*{Acknowledgments}
Part of this work: was supported by the DynBrain project supported by the STICAmSUD international program and was conducted when AQR kept a Ph.D. at Buenos Aires Institute of Technology (ITBA). We would also like to thank Ivana Zorgno for her assistance in the writing.


\bibliographystyle{unsrt}

\end{document}